\documentclass[prl,preprint,showpacs,groupedaddress]{revtex4}
\usepackage{dcolumn}
\usepackage{graphics}
\usepackage{amsmath}
\usepackage{bm}

\usepackage[usenames]{color}

\begin{document}
\title {Photoluminescence from  Microcavities \\ Strongly Coupled to  Single Quantum Dots}
\author{A. Ridolfo$^1$, O. Di Stefano$^1$, S. Portolan$^2$, and S. Savasta$^1$}
\affiliation{$^1$Dipartimento di Fisica della Materia e Ingegneria
Elettronica, Universit\`{a} di Messina Salita Sperone 31, I-98166
Messina, Italy} %\affiliation{$^2$ CNR, Istituto per i
%Processi Chimico-Fisici Sez.\ Messina,Via La Farina 237, I-98123
%Messina, Italy}
\affiliation{$^2$Institute of Theoretical Physics, Ecole
Polytechnique F\'{e}d\'{e}rale de Lausanne EPFL, CH-1015 Lausanne,
Switzerland}
\date{\today}
%\maketitle
%%%%%%%%%%%%%%%%%%%%%%%%%%%%%%%%%%%%%%%%%%%%%%%%%%%%%%%%%
\begin{abstract}

{We study theoretically, the photoluminescence properties of a
single quantum dot in a microcavity under incoherent excitation. We
propose a microscopic quantum statistical approach  providing a
Lindblad (thus completely positive) description of pumping and decay
mechanisms of the quantum dot and of the cavity mode. Our analytical
results show that strong coupling (SC) and linewidths are largely
independent on the pumping intensity (until saturation effects come
into play), in contrast to previous theoretical findings. We shall
show the reliable predicting character of our theoretical framework
in the analysis of various recent experiments.}

\end{abstract}
%\pacs{42.50.Ct, 32.70.Jz, 42.55.Sa, 78.67.Hc}

\maketitle

Cavity quantum electrodynamics (CQED) studies the interaction
between a quantum emitter and a single radiation-field mode. When an
atom is strongly coupled to a cavity mode \cite{atoms}, it is
possible to realize important quantum information processing tasks,
such as controlled coherent coupling and entanglement of
distinguishable quantum systems \cite{Ima}. In this respect
solid-state devices, and in particular semiconductor quantum dots
(QD), are the most promising architectures for the possibility of
miniaturization, electrical injection, control and scalability.
Indeed semiconductor quantum dots provide nanoscale electronic
confinement resulting in discrete energy levels, and an atom-like
light-matter interaction. Cavity quantum electrodynamics addresses
properties of atomlike emitters in cavities and can be divided into
a weak and a strong coupling regimes. For weak coupling, the
spontaneous emission can be enhanced or reduced compared with its
vacuum level by tuning discrete cavity modes in and out of resonance
with the emitter. However, when the interaction strength overcomes
losses, the system enters the so-called strong coupling (SC) regime
 \cite{SC,Reith}. In this case the usual irreversible
spontaneous emission dynamics changes into a reversible exchange of
energy between the emitter and the cavity mode.

Commonly, these solid state systems in the SC regime are
characterized with respect to their behaviour under optical
incoherent pump excitation \cite{Ima,SC,Reith,Arakawa,Finley}.
In this situation, the pump creates an incoherent
population of electron-hole pairs typically far above resonance
which relaxes incoherently into the QD through various different
mechanisms before being emitted by recombination. Moreover, some
experimental observations of SC of QDs in microcavities, indicate
that the cavity mode is weakly coupled to others various electronic
transitions of the system which would contribute in feeding the
cavity mode as well, thus resulting in a second incoherent pumping
channel. The specific interplay of photon and exciton pumping and
decay results in a mixed quantum steady state that influences
considerably the observed spectra as we are going to show in detail
(see Fig.\ 1b). An analogous incoherent excitation would also be
achieved in the case of current injection, highly desired for the
development of optoelectronic quantum devices. Hence, appropriate
theoretical modeling of SC with semiconductor QDs under incoherent
excitation with a reliable predicting character are sought in order
to fulfill the great expectations nowadays attended from future
implementations of SC in QD systems. Of particular interest is the
analysis for increasing pump excitation while the system is brought
into the nonlinear regime.
%The balance among feeding, emission
%and relaxation determines the steady state situation to be analyzed.
%

Recently, Laussy {\em et al.} \cite{Tej PRL} reports on the first
theoretical analysis of this situation. Their master equation
includes pumping as absorption from incoherently populated
reservoirs and spontaneous emission of excitons (radiative
recombination) as well as of photons (cavity losses). Their bosonic
model is well suited for the investigation of the very low
excitation regime where linear optics dominates. It is currently
 heavily exploited (see e.g. Refs. \cite{Arakawa,Finley}) to analyze the SC of single QDs as it, at least at first sight,  displays an impressive
predicting character maintaining a relative easy picture and
analytical results. Nevertheless the results obtained within their
approach display some puzzling features which deserve careful
investigations. As instance, the dependence of the polariton
broadenings, of the Rabi splitting and of the spectra on the
incoherent pumping rates as well as amplification effects are
commonly unexpected results of an harmonic linear dynamics.

\begin{figure}[!ht]
\begin{center}
\resizebox{!}{!}{\includegraphics{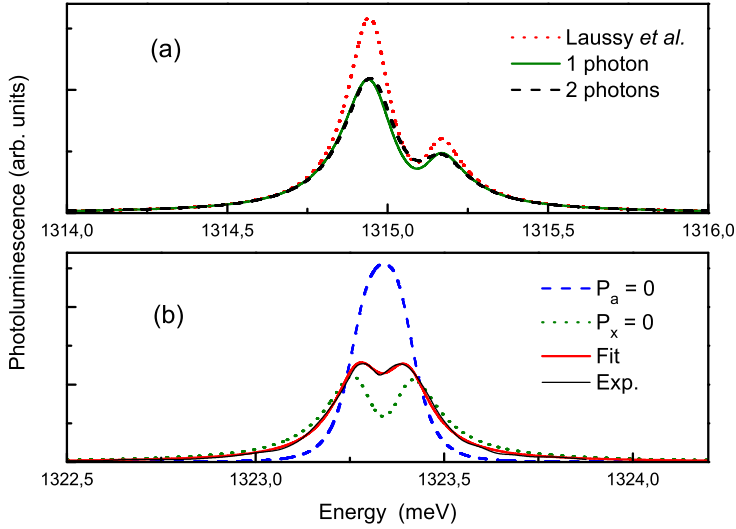}}\caption{(Color online).
(a): PL spectrum
(continuous line) calculated from  Eq.\ (\ref{Spectrum_B}). We used
$\Delta \equiv \omega_a - \omega_x =  - 0.32$ meV, $g =90$ $\mu$eV,
$\gamma_a = 176$ $\mu$eV, $\gamma_x = 133$ $\mu$eV, $P_a = 0.25\,
\gamma_a$, $P_x = 5.3 \cdot 10^{-3}\, \gamma_x$. At these quite low
pump intensities the Bose-like spectrum (\ref{Spectrum_B}) differs
only slightly from that calculated including 2-photon states (dashed
line). The dotted line (red) emission spectrum calculated according to Ref.\
\cite{Tej PRL} differs significantly. (b):
PL spectrum (thin continuous line) from the data reported by Reithmaier {\em et al.}
together with the corresponding fit (red). Dashed and dotted lines (blue and green): PL spectra  with the same parameters as the fit but with $P_a = 0$
and  $P_x = 0$ respectively.
}\label{figura1}.
\end{center}
\end{figure}

\begin{figure}[!ht]
\begin{center}
\resizebox{10.0cm}{!}{\includegraphics{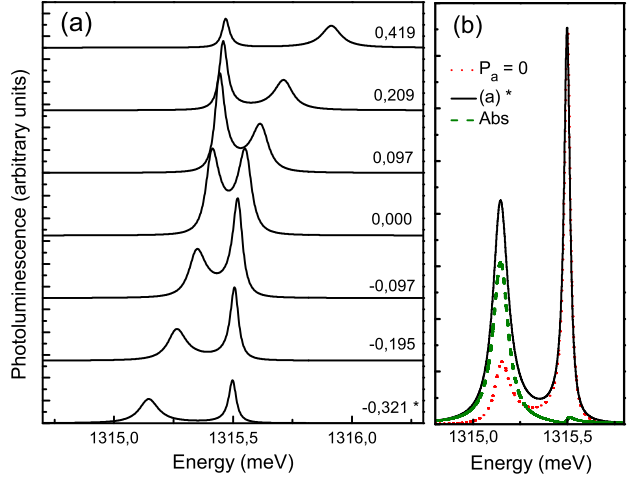}}\caption{Color online. (a) PL spectra
calculated from  Eq.\ (\ref{Spectrum_B} at different detunings reproducing the experimental observation of the SC of a QD-cavity
system reported in Ref.\ \cite{Ima})\label{figura2}. (b) continuous line: one spectrum of panel a (the one with the lowest cavity energy)
compared with the corresponding spectrum obtained using $P_a =0$ (dotted line) and with the corresponding absorption spectrum (dashed line).}

\end{center}
\end{figure}

The authors have the merit to have pointed out
clearly the importance of the effective feeding the other electronic
transitions (viz. the other QDs existing in the sample), weakly
coupled to the cavity mode. This scenario sets a definite departure
from atom-QED models with expected new peculiar features (see e.g. Fig.\
\ref{figura1}b and Fig.\ \ref{figura2}b) which are expected to influence even the
quantum statistics of the emitted quanta. Anyway, although having a
clear and reasonable physical meaning, the master equation of Ref.
\cite{Tej PRL}  is not of Lindblad form and then there is no
guarantee that it safely generates a completely positive (CP) open
dynamics \cite{Lindblad-nota}.
%
%It is worth noticing that the master equation of
%Ref. \cite{Tej PRL}, although having a clear and reasonable physical
%meaning,  is not of Lindblad form and then there is no guarantee
%that it safely generates a completely positive (CP) open dynamics
%\cite{Lindblad-nota}.
%
Actually the obtained photon and exciton populations (see e.g. Eq.\
(3) of Ref. \cite{Tej PRL}) can diverge as soon as pumping rates
approaches the values of the dampings and even assume negative
values as well as it starts to overcome. This highly undesired
feature puts into question also the results obtained at lower
pumping rates when populations remain finite and positive. The
puzzling features, mentioned above, could be the tail (at lower
excitation densities) of such problems. Moreover, the need for a
fresh, flexible and direct tool for the analysis of state-of-the-art
experiments \cite{Arakawa,Finley} is leading various groups to take
advantage of this machinery. These possible artifacts could affect
strongly the interpretations of the experimental data and then have
negative repercussions on applications and device modeling.
%It will become apparent, from our analysis, that the
%principal cause of this unphysical pathology settles in the
%intertwined relationship between absorption and emission within the
%same microscopic dissipative channel (Dire meglio)\cite{Lindblad-nota2}.}

In this letter we provide a theoretical model able to describe SC of a single QD in
a semiconductor microcavity under incoherent excitation in the low
and intermediate excitation regimes where nonlinear optical effects
start to appear. In order to avoid inconsistencies and artifacts, we will start
from a microscopic (though simple) description of the pumping and
decay mechanisms of the QD and the cavity mode.
%=================================================
%
The picture that we have in mind is that of two strongly coupled
subsystems (the cavity mode and the quantum dot excitations), each
in interaction with two independent reservoirs providing both
damping and pumping mechanisms. The laser generates electron-hole
pairs in the continuum wetting layer which, subsequently, relax into
the dot (the $P_x$ pump contribution) by means of incoherent
scattering mechanisms such as exciton-exciton scattering mediated by
phonons. At the same time, the cavity mode is weakly coupled to
others various electronic transitions of the system which would
contribute in feeding the cavity mode as well, resulting in a second
incoherent pumping channel ($P_a$). In the limit of weak excitation
density the resulting dynamics coincides with that of two strongly
coupled harmonic oscillators in the presence of reservoirs.
Nevertheless the obtained analytical results significantly differs from those
of Ref.\ \cite{Tej PRL}. In particular, in the low excitation regime
we obtain a very simple analytical expression for the PL spectrum
which is definite positive, do not display any amplification effect
as well as any change of the Rabi splitting and of the broadenings
as a function of the pump intensities. For higher pump densities
saturation effects and even lasing effects come into play.

%=============================================================

The master equation for this strongly interacting system can be
written as
\begin{equation}\label{model}
    \dot{\rho} = i [\rho, H_S ] + {\cal L}^R_{MC} + {\cal L}^R_{QD}\, ,
\end{equation}
where the system Hamiltonian reads
\begin{equation}
    H_S = \omega_a a^\dag a + \omega_x\, \sigma_+ \sigma_- + g(a^\dag\,  \sigma_- + a\,  \sigma_+)\, ,
\end{equation}
with $g$ being the interaction strength between the cavity mode
(with annihilation operator $a$) at energy $\omega_a$ and the lowest
energy ($\omega_x$) quantum dot exciton with transition operator
from the ground state to the exciton level $\sigma_+ = \left| e
\right>\left< g \right|$ ($\sigma_- = \sigma_+^\dag$). The
superoperators ${\cal L}^R_{MC}$ and ${\cal L}^R_{QD}$ describe the
interaction of the cavity mode and of the QD with the reservoirs
providing both damping and pumping mechanisms.
%============================================================
%
%We model the cavity-mode open dynamics as interacting with two
%reservoirs.
Transmission and diffraction losses of the cavity mode can be
modeled, within a quasimode picture, as an effective coupling
($\gamma^c$) with an ensemble of electromagnetic modes through the
output mirror \cite{Walls,Scully}. When dealing with optical
frequencies it can safely be regarded as a zero temperature
reservoir \cite{Scully}. The second mechanism describes the
incoherent optical pumping of the cavity. It takes into account that
in these samples there may be QDs or more generally electronic
transitions weakly coupled to the cavity, in addition to the one
that undergoes SC, determining an effective pumping of the cavity
mode ($P_a$). By applying the usual Born-Markov and rotating-wave
approximations, we obtain,
\begin{equation}\label{CP cavity}
    {\cal L}^R_{MC} = \frac{P_a + \gamma_a}{2} (2 a \rho a^\dag - a^\dag a \rho - \rho a^\dag a)
    + \frac{P_a}{2} (2 a^\dag \rho a - a a^\dag \rho - \rho a a^\dag)\, ,
\end{equation}
where $\gamma_a = \gamma^c + \gamma^p$ contains contributions from
both the reservoirs and $P_a$ is the total pumping rate depending on
the populations of the electronic levels weakly coupled to the
cavity mode. Assuming only direct and weakly pumped electronic
transitions we obtain $P_a = \sum_i \gamma^p_{i} \langle n_{i}
\rangle$ and $\gamma^p = \sum_i \gamma^p_i$, where $\langle n_{i}
\rangle$ is the population of the $i$-th level and  $\gamma^p_i = 4
\gamma_i g_i^2 /[\gamma_i^2+(\omega_i -\omega_a)^2]$, being
$\gamma_i$ the inverse of the dephasing time of the $i$-th
transition. The form of the two terms in eq.\, (\ref{CP cavity})
clearly shows the physical emerging picture different from that
adressed in Ref.\ \cite{Tej PRL}. Indeed, the cavity experiences
radiative spontaneous emission guided by vacuum fluctuations
(proportional to $\gamma^c$), while the second reservoir provides
both emission (proportional to $\gamma^p_{i} (\langle n_{i} \rangle
+ 1)$) and absorption (to $\gamma^p_{i} \langle n_{i} \rangle$).

The material excitation strongly coupled to the cavity mode is also
under the influence of two different reservoirs: ${\cal L}^R_{QD} =
{\cal L}^{se}_{QD} +    {\cal L}^{P}_{QD}$. The first term describes
spontaneous emission in all the available light modes except the
cavity one,
\begin{equation}
    {\cal L}^{se}_{QD} = (\gamma_x/2)(2 \sigma_- \rho \sigma_+ - \sigma_+\sigma_- \rho - \rho  \sigma_+\sigma_-)\,
    .
\end{equation}
The latter
\begin{equation}
    {\cal L}^P_{QD} = \sum_j (\alpha_j/2)(2 \sigma_{ej} \rho \sigma_{je} - \sigma_{jj} \rho - \rho  \sigma_{jj})\, ,
\end{equation}
is responsible for the incoherent excitation of the QD (e.g. via
phonon induced relaxations) from the $j$-th levels at higher energy
that get populated by optical pumping. The $\alpha_j$ are temperature-dependent phonon-assisted scattering rates from the $j$-th excitonic levels to the  state $\left|e \right>$  \cite{Flagg}.
The transition from the
lowest exciton state $\left| e \right >$ towards higher energy
states can be safely neglected since the experiments of interests
are carried out at $K T << \omega_j -\omega_x$.

This master equation induces an open hierarchy of dynamical
equations. In the SC regime the only meaningful truncation scheme is
the one based on the smallness of the excitation density which
allows to truncate with respect to the number of photon
number-states to be included. The inclusion of one-photon states
only gives linear optical Bose-like dynamics. On the other hand,
once also 2-photon states are taken into account, saturation and
lowest order ($\chi^{(3)}$) nonlinear optical effects arise. Within
our model we are able to go beyond 2-photon nonlinear dynamics. We
shall show essentially non-perturbative calculations, in the sense
that with respect to the chose pumping intensity, we shall always
check the convergence of the presented results to higher-order
number-state truncation, see e.g. Fig.\ \ref{figura1}a.

In the weak excitation regime (linear dynamics) ($\langle \sigma_{ee} \rangle <<
\langle \sigma_{gg} \rangle \simeq 1$), from $\partial_t \left< A
\right > = \text{Tr}(A \dot \rho)$, we obtain
\begin{eqnarray}
    \partial_t \left< a \right> = -i \tilde\omega_a \left< a \right> -i g \left< \sigma_- \right>  \nonumber \\
        \partial_t \left< \sigma_- \right> = -i \tilde \omega_x \left< \sigma_- \right> -i g \left< a \right>\, ,
\end{eqnarray}
which is a system of two coupled damped oscillators in the absence
of any external driving, $\tilde \omega_c = \omega_c - i \gamma_c/2$
with $c=a,x$ . We also observe that the dephasing rates do not
depend on the pumping populations $\langle n_i \rangle$ in contrast
to the results of Ref.\ \cite{Tej PRL}.
%
%
%
%===================================================================================
%===================================================================================
%
We are interested in calculating the  steady-state emission
spectrum: $S(\omega) = \lim_{t \to \infty} 2 \text{Re} \int_0^\infty
\left<a^\dag(t) a(t + \tau) \right> e^{i \omega \tau} d \tau$.
According to the quantum regression theorem, two-time correlations
$\left< A_n(t) A_m(t + \tau) \right>$ follow the same dynamics of
one-body correlation functions $\left<A_m(\tau) \right>$ but with
the one-time correlation $\left< A_n(t) A_m(t) \right>$ as initial
conditions. In our specific case the initial conditions are provided
by the steady-state cavity occupation $n_a = \lim_{t \to
\infty}\langle a^\dag a \rangle$ and by ${\cal C}= \lim_{t \to
\infty}\langle a^\dag \sigma_+ \rangle$, We obtain,
\begin{equation}\label{na}
    n_a = \frac{P_a}{\gamma_a} +\frac{g^2}{\gamma_a}\; \frac{(\gamma_a + \gamma_x)(\gamma_a P_x + \gamma_x P_a)}
    {g^2(\gamma_a + \gamma_x)^2 +\gamma_a \gamma_x \left| \tilde \omega_a - \tilde \omega_x\right|^2}\, ,
\end{equation}
\begin{equation}
    {\cal C}= \frac{g}{\tilde \omega_a - \tilde \omega_x} (n_a -n_x)\, ,
\end{equation}
where $n_x$ can be obtained from Eq.\ (\ref{na}) just exchanging the labels $a$ and $x$, and $P_x = \sum_i
\alpha_i \langle n_{i} \rangle$.
We end up with the following very simple expression, for the PL
spectrum
\begin{equation}\label{Spectrum_B}
    S(\omega) = 2\, \text{Re}\left[\frac{i}{\sqrt{2 \pi}}
    \frac{(\omega - \tilde \omega_x)\, n_a + g\,  {\cal C} }{(\omega- R_1)(\omega -R_2)} \right]\, ,
\end{equation}
where the complex polariton energies determining the spectrum
resonances are given by
\begin{equation}
    R_{(1,2)} = \frac{\tilde \omega_a + \tilde \omega_x}{2} \mp \frac{1}{2}\sqrt{4g^2 +
    (\tilde \omega_a - \tilde \omega_x)^2}\, .
\end{equation}
The analytical structure of eq. (\ref{Spectrum_B})
shows peculiar non-standard features typical of this system, see
Fig.\ \ref{figura1}. Indeed, in addition to the usual dependence in
the denominator on the two Rabi frequencies -- responsible for the
presence of the double-peak structure -- we can appreciate a
numerator carrying a resonant term proportional to the difference
between photon and exciton occupations plus another term
proportional to the sole photon steady-state density $n_a$. Both
terms modulate the Rabi resonances with respect to the pumping
scenario they depend upon. {\em According to Eq.\ (\ref{Spectrum_B}) it is not possible that a system in the weak coupling regime enters
SC thanks to pumping in  contrast to the results of Ref.\ \cite{Tej PRL}}.

Although at low pump intensities, our approach and that of Ref.\
\cite{Tej PRL} essentially represent models of a linear Bose-like
dynamics of two coupled harmonic oscillators, nontrivial differences
can be appreciated. Indeed, Fig.\ \ref{figura1}a displays the PL
spectrum (continuous line) calculated from  Eq.\ (\ref{Spectrum_B})
compared with the result obtained from the model of Ref.\ \cite{Tej
PRL} for the same situation (see caption of Fig.\ \ref{figura1} for
details). At these quite low pump intensities our Bose-like spectrum
(\ref{Spectrum_B}) differs only slightly from that calculated
including 2-photon states (dashed line) and the inclusion of
additional photon states does not produce appreciable modifications
of the spectrum. On the contrary the emission spectrum calculated
according to Ref.\ \cite{Tej PRL} differs significantly (dash-dotted
line). The difference originates mainly from the effective reduction
of the damping rates due to the incoherent pump \cite{Tej PRL}, a
feature absent in our theoretical description. The model presented
in Ref.\ \cite{Tej PRL} reproduces the emission spectra measured by
Reithmaier {\em et al.} \cite{Reith}. Although the resulting fit is
in excellent agreement with the experimental data, some key fitted
parameters (e.g. the pumping rates) appears to be in contrast with
the experimental results. In particular, as can be inferred from
Figs.\ 3 and 4 of Ref.\ \cite{Reith}, the measured PL intensities
display a very small variation as a function of temperature. On the
contrary they estimate pumping rates with a variation larger than a
factor four \cite{Tej2}. In addition, Reithmeier {\em et al.}
provides a quite accurate estimate of the mean cavity photon-number
(see Additional Information of Ref.\ \cite{Reith}) which is one
order of magnitude lower than the result of the above montioned fit.
This disagreement strongly supports the absence of any dependence of
the damping rates on the pumping rates in agreement with the results
of our model. Indeed, all the above experimental
features are well reproduced by our master equation (\ref{model}).

Despite of its analytical simple form, eq. (\ref{Spectrum_B}) shows
a nontrivial dependence of the system on the ratio between the
cavity and dot pumping rates. Fig.\ \ref{figura1}b displays one PL
spectrum (thin continuous line) from the data reported by Reithmaier
{\em et al.} together with the corresponding fit we obtain using
Eq.\ (\ref{Spectrum_B}) (thicker continuous line). Beside the nearly
coincidence of the two curves in this situation, Eq.\
(\ref{Spectrum_B}) is also in very good agreement with the
experimental spectra of Ref. \cite{Reith} at different detunings. As
absolute information on the PL intensities is commonly not available
from this kind of experiments, we cannot obtain absolute values for
the pumping rates. Instead we gather from the fit the ratio $P_a/P_x
= 0.86$. We also obtain $g = 76$ meV, $\gamma_a = 100$ $\mu$eV,
$\gamma_x = 35$ $\mu$eV. In order to evidence the impact that the
pumping mechanism can have on SC emission lines even at very small
detuning, we plotted in Fig.\ \ref{figura1}b two spectra with the
same parameters as the fit but with $P_x = 0$ (dotted line) and $P_a
= 0$ (dashed line). These plots show that, close to the SC
threshold, incoherent pumping of the exciton level can even hinder
the line splitting. This is the case of the system investigated by
Reithmaier {\em et al.} \cite{Reith}. This result put forward the
importance of adressing the correct mixed quantum steady state in
order to describe the experimental results.
%
%
%
%===================================================================================
%===================================================================================
%

Another situation for a similar system has been recently reported in
Ref.\ \cite{Ima}. Some first results based on the present
multi-photon model (\ref{model}) are depicted in Fig.\
\ref{figura2}a. It displays emission spectra at different detunings
reproducing one experimental observation of the SC of a QD-cavity
system reported in Ref.\ \cite{Ima}. Although the parameters have
been chosen in order to fit just one of the spectra, our model
provides results which agree very well \cite{nota} with all the
spectra presented in Fig.\ (2b) of Ref. \cite{Ima} changing only the
cavity energy. We can thus unravel the ratio between the cavity-mode
and the exciton pumping in this kind of experiments. In this case we
obtain $P_a/P_b =0.13$. Fig.\ \ref{figura2}b
clearly underlines the importance of taking into account the correct
quantum steady state resulting from the interplay of pumping and
decay (of both the cavity mode and the exciton level) in order to
describe properly the SC experiments.
%The importance of taking into
%account the correct quantum steady state resulting from the
%interplay of pumping and decay (of both the cavity mode and the
%exciton level), in order to describe properly the SC experiments,
%can be further appreciated in Fig.\ \ref{figura2}b.
It shows one spectrum of Fig.\ \ref{figura2}a
($\Delta = 0.1$ meV) (continuous line) together with the emission
spectrum obtained neglecting the pumping $P_a$ of the cavity mode
and the absorption spectrum under a coherent pump. As can be clearly
seen without the proper $P_a$ pumping mechanism, the calculated
results differ significantly from the experiment.
%
%
%
%
%
%===================================================================================
\begin{figure}[!ht]
\begin{center}
\resizebox{!}{!}{\includegraphics{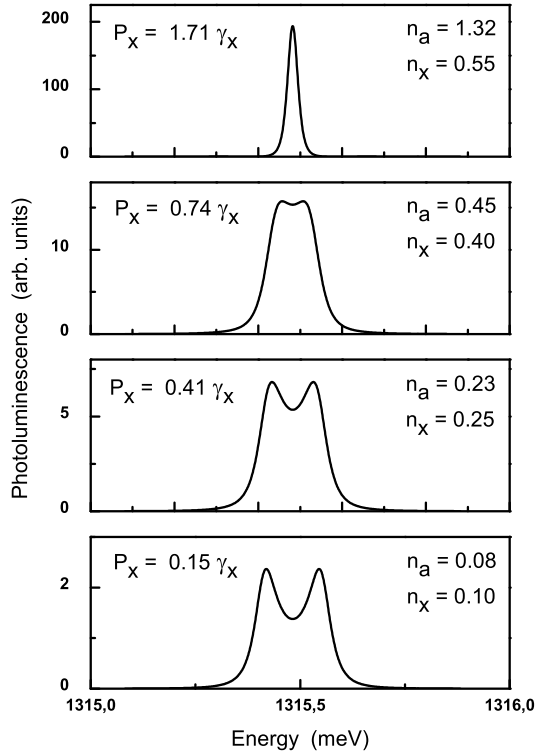}}\caption{Transition from Rabi splitting to
 single-emitter laser when increasing the pumping rate.}\label{figura3}.
\end{center}
\end{figure}
%===================================================================================
%

Our model is also well suited to describe situations well beyond
that of linear or lowest order nonlinear effects. Fig.\
\ref{figura3} displays emission spectra obtained at different
pumping intensities. These results have been obtained after an exact
calculation of the SC quantum dynamics achieved truncating the
photon number-states only when including larger numbers (and hence
dealing with a larger system of equations of motion) does not
produce any change in the emission spectrum. We used the same
broadenings  of Fig.\ \ref{figura2}, a coupling $g = 76$ $\mu$eV and
we set $P_a =0$ and $\Delta = 0$. Increasing the excitation a
reduction of the Rabi splitting due to saturation effects is clearly
observable. The upper panel show that as soon as a small popoulation
inversion ($n_x > 0.5$) is reached an important narrowing of the
linewidth appears indicating a thresholdless lasing behaviour.
It seems worth underlining that, within our model,
both the linewidth narrowing and the build up of lasing are purely
fermionic effects not to be confused with the  narrowing due
to the incoherent pumping found in the model of Ref.\ \cite{Tej PRL}
(see also \cite{Arakawa}).

%\newpage

In conclusion, we have given a simple theoretical framework with a
reliable predicting character for the analysis of photoluminescence
properties under incoherent excitation of single quantum dot
microcavity devices in a steady state maintained by a continuous
incoherent pumping. On the contrary to currently exploited
approaches in the literature, we constructed a completely positive
quantum master equation able to provide physically sensible results
free from artifacts do to the reduced phenomenological description.
Remarkably, our theoretical framework represents a simple and
flexible tool naturally suitable for the analysis of the impact of
the interplay between the two feeding mechanisms (i.e. $P_a$ and
$P_x$) onto the quantum statistic of the emitted quanta at different
excitation intensities (under current development). At low pumping
rates we obtained an analytical expression for the emission spectrum
which directly shows the very different impact of the two feeding
mechanisms on the spectra. Our model showed a great flexibility
proposing simple but sensible descriptions of various experiments in
the strong-coupling regime. It provides essentially non-perturbative
results which can be exploited for the efficient modeling of future
SC optoelectronic devices.

We thanks Fausto Rossi for helpful suggestions and discussions
on Lindblad master equations and  completely positive open
dynamics.
%-------------------------------------------------------------------
%\begin{references}


\begin{thebibliography}{50}
\bibitem{nota} Our calculated spectra are in very good agreement  with those of Ref.
\cite{Ima} except for the central peak which up to now escapes
a precise physical explanation. This feature is present at every
detuning and it is specific to the investigated sample.
\end{thebibliography}
\end{document}